\documentclass[11pt,a4paper]{article}
\usepackage{amsmath}
\usepackage{amsfonts,amssymb}
\begin{document}

\newcommand{\nl}{\nonumber\\}
\newcommand{\nnl}{\nl[6mm]}
\newcommand{\nle}{\nl[-2.3mm]\\[-2.3mm]}
\newcommand{\nlb}[1]{\nl[-2.3mm]\label{#1}\\[-2.3mm]}

\newcommand{\be}{\bes}
\newcommand{\ee}{\ees}
\newcommand{\bes}{\begin{eqnarray}}
\newcommand{\ees}{\end{eqnarray}}
\newcommand{\eens}{\nonumber\end{eqnarray}}

\renewcommand{\/}{\over}
\renewcommand{\d}{\partial}

\newcommand{\Np}[1]{{N+p\choose N #1}}
\newcommand{\Npr}{{N+p-r\choose N-r}}

\newcommand{\eps}{\epsilon}
\newcommand{\dlt}{\delta}
\newcommand{\rep}{\varrho}

\newcommand{\mm}{{\bf m}}
\newcommand{\nn}{{\bf n}}
\newcommand{\phim}{\phi_\mm}
\newcommand{\phin}{\phi_\nn}

\newcommand{\dmu}{\d_\mu}
\newcommand{\dnu}{\d_\nu}
\newcommand{\qmu}{q^\mu}
\newcommand{\xmu}{\xi^\mu}

\renewcommand{\L}{{\cal L}}
\newcommand{\Lxi}{\L_\xi}

\newcommand{\oj}{{\mathfrak g}}
\renewcommand{\pm}{(-1)^{\eps(\rep)}}

\newcommand{\vect}{{\mathfrak{vect}}}
\newcommand{\ZZ}{{\mathbb Z}}
\newcommand{\TT}{{\mathbb T}}

\title{{Resolution of the problem of time in quantum gravity}}

\author{T. A. Larsson \\
Vanadisv\"agen 29, S-113 23 Stockholm, Sweden,\\
 email: Thomas.Larsson@hdd.se }

\maketitle
\begin{abstract}
The metric determines the casual structure of spacetime, but in 
quantum gravity it is also a dynamical field which must be quantized
using this causal structure; this is the famous problem of time.
A radical resolution of this paradox is
proposed: remove the concept of space-like separation entirely. This
can be done by describing all fields in terms on $p$-jets, living on
the observer's trajectory; all points on the trajectory have 
time-like separations. Such a description is necessary to construct
well-defined representations the $N$-dimensional generalization of
the Virasoro algebra $Vir(N)$; this is the natural quantum extension
of $\vect(N)$, which is the correct symmetry algebra of general 
relativity in $N$ dimensions. The limit
$p\to\infty$, necessary for a field theory interpretation,
only exists if $N=4$ and there are three fermions for every two bosons,
a relation that is satisfied in the standard model coupled to gravity.
\end{abstract}

\newpage
All known physics can be well described by general relativity,
quantum theory, and the standard model. Unfortunately, the first two
theories are known to be mutually inconsistent. Since no
conclusive experimental hints of additional physics have been seen, it
is reasonable to expect that the ultimate theory should be very close
to the above theories. During the crisis in the 1930s, when it was 
realized that second-order perturbation theory for QED is divergent, 
many people suggested that radically new physics was needed, but the
resolution turned out to be much more mundane. Renormalization can be
viewed as a mere reinterpretation, albeit a radical one, of 
na\"\i ve QFT.

The situation today appears quite similar. There are several 
formulations of quantum theory, which produce the same results in
flat space, but none of these can handle gravity. However, a more
fundamental formulation of quantum physics could exist, which both
is consistent with gravity and reduces to the known theories in flat 
space. That the interpretation of quantum theory is still debated 
75 years after its inception can be taken as a hint that standard
formulations may not be self-evident. In contrast, the interpretation
of general relativity is seldom debated.

The only viable way to find such a formulation is to consider the 
correct symmetries of quantum gravity. The correct symmetry of 
classical gravity, i.e. general relativity, is the full spacetime
diffeomorphism group (without background fields), supplemented by local 
Lorentz transformations when spinors are needed. 
However, in quantum physics symmetries are only represented projectively;
we are thus led to projective representations of the spacetime 
diffeomorphism group. On the Lie algebra level, this means
that the algebra of vector fields in $N$ spacetime dimensions,
$\vect(N)$, acquires an abelian extension\footnote{In previous writings 
I have called this algebra the diffeomorphism algebra $diff(N)$.
Note that the Virasoro-like extension of $\vect(N)$ arises from
projective representations of spacetime diffeomorphisms, not from
some membrane generalization of Weyl scalings in string theory.}:
the $N$-dimensional Virasoro algebra $Vir(N)$. 

The construction of projective
representations of $\vect(N)$ is not trivial when $N>1$. The usual
prescription, namely to introduce canonical momenta and normal order,
does not work. One way to see this is that such a recipe would 
produce a central extension, but the Virasoro-like cocycle is 
non-central when $N>1$. The route around this problem was first found
by Rao and Moody \cite{RM94} in 1994, and explained 
geometrically by myself \cite{Lar98}. 
The statement that quantum theory and gravity are incompatible is 
thus no longer compeletely true; since 1994, a consistent marriage 
between quantization and general covariance exists on the symmetry level.

The crucial idea is to first expand all 
fields in a multi-dimensional Taylor series around the points of a 
one-dimensional curve (the observer's trajectory $\qmu(t)$),
\be
\phi(x) = \sum_{|\mm|\leq p} {1\/\mm!} \phim(t) (x-q(t))^\mm,
\label{Taylor}
\ee
where $\mm$ is a multi-index and $|\mm|$ denotes its length. Standard
multi-index notation is used. Such objects transform as
\bes
[\Lxi, \phim(t)] &=& -\sum_{|\nn|\leq|\mm|} T^\nn_\mm(\xi(q(t))) \phin(t), 
\nlb{Freal}
{[}\Lxi, \qmu(t)] &=& \xmu(q(t)).
\eens
Explicit expressions for the matrices $T^\nn_\mm(\xi)$ are given in 
\cite{Lar98}. By truncation to $|\mm|\leq p$, we obtain a realization of 
$\vect(N)$ on the space of trajectories in the space of tensor-valued
$p$-jets. This space consists of finitely many functions of a single
variable, which is precisely the situation where the normal ordering
prescription works. 

To make the connection to the Virasoro algebra very explicit, it is
instructive to write down the brackets in a Fourier basis. Start with
the Virasoro algebra $Vir$:
\be
[L_m, L_n] = (n-m)L_{m+n} - {c\/12} (m^3-m) \delta_{m+n},
\ee
where $\delta_m$ is the Kronecker delta. When $c=0$, 
$L_m = -i \exp(imx) d/dx$, $m \in \ZZ$. 
The element $c$ is central, meaning that it commutes with all of $Vir$;
by Schur's lemma, it can therefore be considered as a c-number.
Now rewrite $Vir$ as
\bes
[L_m, L_n] &=& (n-m)L_{m+n} + c m^2 n S_{m+n}, \nl
{[}L_m, S_n] &=& (n+m)S_{m+n}, \nle
{[}S_m, S_n] &=& 0, \nl
m S_m &=& 0.
\eens
It is easy to see that the two formulations of $Vir$ are equivalent
(I have absorbed the linear cocycle into a redefinition of $L_0$).
The second formulation immediately generalizes to $N$ dimensions.
The generators are $L_\mu(m) = -i \exp(i m_\rho x^\rho) \dmu$ and
$S^\mu(m)$, where $x = (x^\mu)$, $\mu = 1, 2, ..., N$ is a point in 
$N$-dimensional space and $m = (m_\mu)$. The Einstein convention is used;
repeated indices, one up and one down, are implicitly summed over. 
The defining relations are
\bes
[L_\mu(m), L_\nu(n)] &=& n_\mu L_\nu(m+n) - m_\nu L_\mu(m+n) \nl 
&&  + (c_1 m_\nu n_\mu + c_2 m_\mu n_\nu) m_\rho S^\rho(m+n), \nl
{[}L_\mu(m), S^\nu(n)] &=& n_\mu S^\nu(m+n)
 + \delta^\nu_\mu m_\rho S^\rho(m+n), \nle
{[}S^\mu(m), S^\nu(n)] &=& 0, \nl
m_\mu S^\mu(m) &=& 0.
\eens
This is an extension of $\vect(N)$ by the abelian ideal with basis 
$S^\mu(m)$. This algebra is even valid globally on the 
$N$-dimensional torus $\TT^N$.
Geometrically, we can think of $L_\mu(m)$ as a vector field 
and $S^\mu(m) = \eps^{\mu\nu_2..\nu_N}S_{\nu_2..\nu_N}(m)$ 
as a dual one-form (and $S_{\nu_2..\nu_N}(m)$ as an $(N-1)$-form);
the last condition expresses closedness. 

The cocycle proportional to $c_1$ was discovered by 
Rao and Moody \cite{RM94}, and the one proportional to $c_2$ by
myself \cite{Lar91}. There is also a similar multi-dimensional 
generalization of affine Kac-Moody algebras,
presumably first written down by Kassel.
The multi-dimensional Virasoro and 
affine algebras are often refered to as ``Toroidal Lie algebras''
in the mathematics literature. 

One may think of the projective representation theory of $\vect(4)$
as the kinematics of quantum gravity. The kinematics of special
relativity is governed by the Poincar\'e algebra, with the result 
that any special-relativistic theory must be formulated in terms of
Minkowski tensor and spinor fields. Similarly, general covariance 
dictates that general relativity must use only $\vect(4)$ modules,
such as tensor densities, connections, and closed forms. Even if 
dynamics, e.g. Einstein's equation, does not follow from symmetry
considerations, understanding kinematics is an important first step, 
and many properties of a theory follow from kinematics alone.
In particular, there are two important properties of quantum gravity 
that follow directly from kinematics: resolution of the
causality paradox and the existence of both bosons and fermions.

Now let us discuss the main topic of this note, namely the problem of
time (referred to as the ``casuality paradox'' in a previous version 
of this e-print). On the one hand, the metric
determines the causal structure of spacetime (time-like, space-like,
light-like). On the other hand, in gravity the metric is a dynamical field
and therefore it needs to be quantized using the very same 
causal structure. This is definitely a problem in the Hamiltonian 
formulation, which involves a manifest split into space and time, but
it also appears in disguised form in the path integral formalism. 

The resolution forced upon us by representation theory is as radical 
as it is simple: do not consider space-like separations at all. After
all, a quantum theory should only deal directly with observable 
quantities, and space-like distances can not be observed; no observer
can be in two different places at the same time. To introduce two
observers does not help, because the second observer belongs to the 
system being observed by the first observer. Of course, I do not 
propose that space-like distances do not exist, only that they are
not described explicitly within the formalism. 
What I {\em do} propose is a
very strong notion of locality. Not only should all interactions be
local in spacetime, but the theory should only deal directly with
quantities that are local to the observer, i.e. objects on the 
observer's trajectory. A drastic example: a terrestial observer does 
not observe the sun itself, but only photons and other particles that
reach terrestial detectors, including the naked eye. 

Of course, it must be possible to reconstruct distant events from 
local data. The sun certainly exists, but its existence might not 
a primary observable on earth. Mathematically, 
the restriction to local data amounts to a Taylor expansion 
around the observer's trajectory prior to quantization, as in 
Eq. (\ref{Taylor}). Ultimately, we must reconstruct the quantum field 
$\phi(x)$ throughout spacetime. Classically, this can be done using
the same Taylor expansion. The existence of a $p\to\infty$ limit can thus 
be thought of as a statement of objective reality. Modulo questions of 
convergence, there is 
no problem to take this limit in (\ref{Taylor}), which 
hence determines the field in terms of jet data\footnote{ Eq. 
(\ref{Taylor}) really defines a field $\phi(x,t)$. To ensure 
$t$-independence one must require that $\dot\phi_\mm(t) = 
\dot\qmu(t)\phi_{\mm+\mu}(t)$.}. In the quantum case
there is an important obstruction: the abelian charges, i.e. the 
parameters multiplying the cocycles, diverge with $p$. 

As explained in \cite{Lar98}, the correct algebra to consider is 
really $\vect(N)\oplus\vect(1)$, where the extra $\vect(1)$ factor
describes reparametrizations of the observer's trajectory. After 
normal ordering, this algebra acquires four different Virasoro-like 
cocycles, giving rise to the {\em DRO (Diffeomorphism, 
Reparametrization, Observer) algebra} $DRO(N)$. 
\bes
[\Lxi,\L_\eta] &=& \L_{[\xi,\eta]} 
 + {1\/2\pi i}\int dt\ \dot q^\rho(t) 
 \Big\{ c_1 \d_\rho\dnu\xmu(q(t))\dmu\eta^\nu(q(t)) +\nl
&&\quad+ c_2 \d_\rho\dmu\xmu(q(t))\dnu\eta^\nu(q(t)) \Big\}, \nl
{[}L_f, \Lxi] &=& {c_3\/4\pi i} \int dt\ 
 (\ddot f(t) - i\dot f(t))\dmu\xmu(q(t)), 
\label{DGRO}\\
{[}L_f,L_g] &=& L_{[f,g]} 
 + {c_4\/24\pi i}\int dt (\ddot f(t) \dot g(t) - \dot f(t) g(t)), \nl
{[}\Lxi,\qmu(t)] &=& \xmu(q(t)), \nl
{[}L_f,\qmu(t)] &=& -f(t)\dot\qmu(t),
\eens
where $\xi = \xmu(x)\dmu, \eta \in \vect(N)$ and $f = f(t)d/dt, g \in
\vect(1)$.
If $\phi(x)$ is a tensor field of type $\rep$, the four abelian
charges are given by
\bes
c_1 &=& 1 +\pm k_1(\rep)\Np{} +\pm \dim\,\rep {N+p+1\choose N+2}, \nl
c_2 &=& \pm k_2(\rep)\Np{} +\pm 2k_0(\rep)\Np{+1} \nl
&& +\pm \dim\,\rep {N+p\choose N+2}, 
\\
c_3 &=& 1 - \pm k_0(\rep)\Np{} - \pm \dim\,\rep\Np{+1}, \nl
c_4 &=& 2N+\pm\dim\,\rep \Np{}, 
\eens
where $\eps(\rep)$ is the Grassmann parity of the field associated with
$\rep$, and 
$k_0(\rep)$, $k_1(\rep)$, $k_2(\rep)$, and $\dim\,\rep$ are finite
numbers that depend on the choice of $gl(N)$ representation only.
If $\rep = \oplus_i \rep_i$ is reducible, we must interpret 
$\pm k_0(\rep) = \sum_i (-1)^{\eps(\rep_i)} k_0(\rep_i)$.

We see that all abelian charges diverge; the worst case is 
$c_1 \approx c_2 \approx p^{N+2}/(N+2)!$, which diverge in all 
dimensions $N > -2$. In \cite{Lar01} I devised a way out of this 
dilemma: consider a more general realization on several jets 
$\phim^{(i)}(t)$, like in (\ref{Freal}), but with different values of 
the jet order $p$. Take $r+1$ jets, with $p$ replaced by
$p$, $p-1$, ..., $p-r$, respectively, and with $\rep$ replaced by
$\rep^{(i)}$ in the $i$:th term. If we choose the $gl(N)$ 
representations $\rep^{(i)}$ suitably, it is possible to cancel the 
leading divergencies. For simplicity, I only consider $c_4$ here.
Let $r=4$ and choose $\rep^{(i)}$ such that
\bes
&&\dim\,\rep^{(0)} = x, \qquad
\dim\,\rep^{(1)} = -4x, \qquad
\dim\,\rep^{(2)} = 6x, 
\nlb{dims}
&&\dim\,\rep^{(3)} = -4x, \qquad
\dim\,\rep^{(4)} = x,
\eens
where the sign reflects Grassmann parity. In the sum representation,
the abelian charge $c_4$ becomes
\be
c_4 = 2N+(-1)^{\eps(\rep_0)} x {N+p-4\choose N-4},
\ee
which has a finite non-zero $p\to\infty$ limit in $N=4$ dimensions
only. More generally, it turns out that all abelian charges can be 
made finite with $r+1$ representations $\rep^{(i)}$ when $N=r$.
We see from the signs in (\ref{dims}) that a well-defined 
$p\to\infty$ limit requires both bosons and fermions to be present
in the theory, without the need for supersymmetry. 

It is unavoidable that the multi-dimensional Virasoro algebra arises in
quantum gravity; it is the mathematical expression for quantization of
general covariance, which is the correct symmetry of classical gravity.
Although it can only be said to contain quantum gravity kinematics at
this stage, it is promising that a long-standing conceptual problem such
as the problem of time just disappears. 

Moreover, dynamics can be
introduced using ideas from the antifield approach to gauge theories;
jets $\phim^{(i)}(t)$ of non-maximal order take the role of 
anti-fields rather than genuine fields. 
In combination with the result above and some natural
assumptions about the Euler-Lagrange equations, it leads to the
conclusion that spacetime must have four dimensions and (less
conclusively) that there are three fermions for every two bosons, with
the na\"\i ve counting of degrees of freedom; this relation holds in the
standard model coupled to general relativity \cite{Lar02}.

In both quantum mechanics and general relativity, the observer (or test
particle) plays an important but somewhat peripheral role. The concept
appears in the interpretations of the theories, but not in the core
(Schr\"odinger and Einstein) equations. Now the observer has taken the
final step into the core mathematical formalism.


\begin{thebibliography}{99}

\bibitem{Lar91} T.A. Larsson,
  {\em Central and non-central extensions of multi-graded Lie algebras},
  J. Phys. A. {\bf 25} (1992) 1177--1184 

\bibitem{Lar98} T.A. Larsson,
  {\it Extended diffeomorphism algebras and trajectories in jet space},
  Comm. Math. Phys. {\bf 214} (2000) 469--491 
  {\tt math-ph/9810003}

\bibitem{Lar01} T.A. Larsson, 
  {\it Multi-dimensional diffeomorphism and current algebras from
  Virasoro and Kac-Moody Currents},
  {\tt math-ph/0101007} (2001)

\bibitem{Lar02} T.A. Larsson, 
  {\it Koszul-Tate cohomology as lowest-energy modules of non-centrally
  extended diffeomorphism algebras},
  {\tt math-ph/0210023} (2002)

\bibitem{RM94} S.E. Rao and R.V. Moody,
  {\it Vertex representations for $N$-toroidal Lie algebras and a
  generalization of the Virasoro algebra},
  Comm. Math. Phys. {\bf 159} (1994) 239--264 

\end{thebibliography}
\end{document}